\documentclass[aps,pra,twocolumn,groupedaddress]{revtex4} %submit to arXiv

\usepackage{graphicx}
\usepackage{bm}
\begin{document}

%\preprint{}

%\title{Spectroscopic determination of the $^{86}$Sr and $^{88}$Sr $s$-wave scattering lengths}
\title{Two-photon photoassociative spectroscopy of ultracold $^{88}$Sr}


\author{  Y. N. Martinez de Escobar$^1$,  P. G. Mickelson$^1$, P. Pellegrini$^2$, S. B. Nagel$^1$, A. Traverso$^1$,  M. Yan$^1$, R. C\^ot\'e$^2$ and T. C.
Killian$^1$}
%\author{The Usual Suspects}

%\thanks{}
\affiliation{$^1$Rice University, Department of Physics and
Astronomy, Houston, Texas, 77251}

%\author{P. Pellegrini, and R. C\^ot\'e}

%\thanks{}
\affiliation{$^2$Department of Physics, U-3046, University of
Connecticut, Storrs, CT, 06269-3046}


\date{\today}

\begin{abstract}
We present results from two-photon photoassociative spectroscopy of
the least-bound vibrational level of the X$^1\Sigma_g^+$ state of
the $^{88}$Sr$_2$ dimer. Measurement of the binding energy allows us
to determine the $s$-wave scattering length, $a_{88}=-1.4(6)\,a_0$.
For the intermediate state, we use a bound level on the metastable
$^1S_0$-$^3P_1$ potential, which provides large Franck-Condon
transition factors and narrow one-photon photoassociative lines that
are advantageous for observing quantum-optical effects such as
Autler-Townes resonance splittings.

\end{abstract}

%\keywords{}

\maketitle

%\section{Introduction\label{introduction}}
%\subsection{}
%\subsubsection{}

% If in two-column mode, this environment will change to single-column
% format so that long equations can be displayed. Use
%\begin{widetext}
%\end{widetext}

\section{Introduction\label{introduction}}
%\subsection{}
%\subsubsection{}

%molecular production rate

%weakest bound level ever measured

%emphasize that this is a narrow transition - long-lived excited
%state, low powers

\begin{figure}%%%%%Use Screen size centered on page when exporting from matlab!!!
 %%THis saves the sizes%
%print to pdfwriter from ppoint,open resulting pdf with ghostview, convert to
%eps with pswrite, max resoltuion
  %\includegraphics[width=3in,clip=true,trim=100 290 50 80 ]{energydiagramstraight.eps}\\
  %\includegraphics[width=3in,clip=true]{PASdrawing.eps}\\
  \includegraphics[width=3.75in,clip=true, trim=25 30 0 50, angle=0]{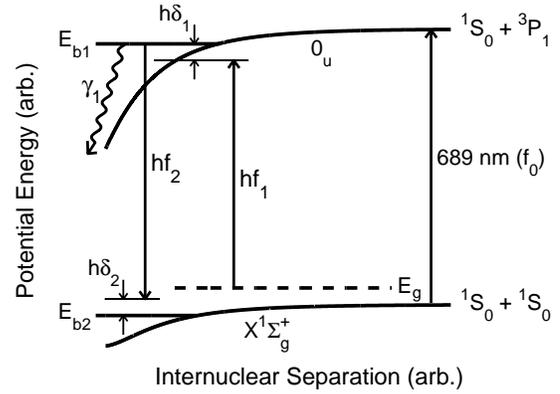}\\ %{node86.eps}\\
  \caption{Two-photon PAS diagram. The energy of two well-separated $^1S_0$ atoms
  at
  rest is taken as zero. $E_g$ is the kinetic energy of the colliding atom pair.
  $E_{b_1}$ is the unperturbed energy of the
  bound state of the excited molecular potential that is near resonance with the
  free-bound laser.
  $E_{b_2}$ ($<0$) is the unperturbed energy of the
  bound state of the ground molecular potential. The photon of
  energy $h f_1$  is detuned from $E_{b1}$ by
  $h\delta_1$, while the photon of
  energy $h f_2$  is detuned from $E_{b2}$ by
  $h\delta_2$. The decay rate of $b_1$ is $\gamma_1$.
  Stark shifts of the levels due to trapping laser
  fields are neglected in this schematic.
  %All frequencies
%are measured with respect to the atomic $^1S_0$-$^1P_1$ transition.
  }\label{PASDiagram}
\end{figure}

Precise knowledge of interactions between ultracold atoms has
enabled spectacular  advances in the production and study of quantum
gases\cite{jtl06}. The most accurate tool for determining those
interactions is spectroscopy of bound molecular states, such as
two-photon photoassociative spectroscopy (PAS) in which two laser
fields couple colliding atoms  to a weakly bound state of the ground
molecular potential via a near-resonant intermediate state (Fig.\
\ref{PASDiagram}). Two-photon PAS has been used to measure binding
energies in Li \cite{amg97,ams95},   Na \cite{vve99}, K
\cite{wne00}, Rb \cite{tfv97}, He \cite{mpk06}, and Yb \cite{kek08}.
Each of these measurements provides accurate determination of the
atomic $s$-wave scattering length  ($a$) and understanding of the
path towards quantum degeneracy and  behavior of resulting quantum
fluids. Here, we report two-photon PAS of $^{88}$Sr and
determination of $a$ for the ground molecular
 potential (X$^1\Sigma_g^+$). Through mass-scaling, we also determine
$a$ for all stable-isotope collisional combinations.

Alkaline-earth atoms such as Sr, and atoms with similar electronic
configuration, differ significantly from alkali-metal atoms that are
typically used in ultracold experiments. They have a closed-shell
ground state structure, numerous isotopes including spinless bosons,
and metastable triplet levels that lead to novel laser-cooling
techniques \cite{kii99} and interactions \cite{ksg03,nsl03}. They
present many new opportunities for the study and application of
ultracold atoms, such as optical frequency standards \cite{lzc08},
long-coherence-time interferometers \cite{fps06}, and Bose and Fermi
quantum degenerate gases and mixtures \cite{tmk03,ftk07}. While the
collisional properties of these atoms have been the subject of
intense study
\cite{ctk04,mjs01,ctj05,nju06,zbl06,mms05,asp03,dbw03}, until now
precise scattering length values have only been published for Yb
\cite{kek08}.

For two-photon PAS of Sr, we utilize an intermediate state that is
bound in the $0_u$ potential that corresponds to the $^1S_0+
{^3P_1}$ asymptote at long range. The spin-forbidden $^1S_0$-$
{^3P_1}$ intercombination transition at $\lambda=689$\,nm is weakly
allowed due to spin-orbit coupling of the ${^3P_1}$ state with the
lowest-lying ${^1P_1}$ level \cite{scg04theory}. PAS involving an
intercombination line \cite{ctk04,zbl06,tke06} transition differs
qualitatively from spectroscopy near an electric dipole-allowed
transition  due to the metastability of the $^3P_1$ state
($\tau=21.5$\,$\mu$s \cite{zbl06}). This increases the importance of
the van der Waals interaction relative to the dipole-dipole term in
determining the shape of the excited molecular potential, which
makes the ground and excited potentials more similar than in
alkali-metal atoms. Resulting Franck-Condon factors for transitions
from free-atom to weakly bound excited molecular states are smaller,
but overlap integrals between excited and ground molecular states
are larger, which has implications for the formation of ground state
molecules. It has been predicted that an optical Feshbach resonance
induced by a laser tuned near an intercombination transition
\cite{fks96} can change the ground state scattering length with much
lower inelastic loss \cite{ctj05} than when using electric
dipole-allowed transitions \cite{fjl00,ttw04}. Long coherence times
are helpful for observing quantum optical effects, such as
Autler-Townes splittings \cite{ato55} of molecular levels
\cite{sds03,tdp01,dwj05}, which is closely related to creation of an
atom-molecule dark state \cite{wtt05,mpk06} and state-selective
production of ultracold ground state molecules
\cite{wfh00,tdp01,rbm04}.

\section{Experimental Setup\label{Section:Experimental Setup}}

%The initial aspects of the experimental setup are similar to those
%described previously [PRL 94,083004 (2005) and PRL 95, 223002
%(2005)], but the addition of a far-off-resonant dipole trap (FORT)
%is new.

 \begin{figure}%%%%%Use Screen size centered on page when exporting from matlab!!!
 %%THis saves the sizes%
%print to pdfwriter from ppoint,open resulting pdf with ghostview, convert to
%eps with pswrite, max resoltuion
  %\includegraphics[width=3in,clip=true,trim=100 290 50 80 ]{energydiagramstraight.eps}\\
  %\includegraphics[width=3in,clip=true]{PASdrawing.eps}\\
\includegraphics[width=2.25in,clip=true, trim=40 20 220 0, angle=270]{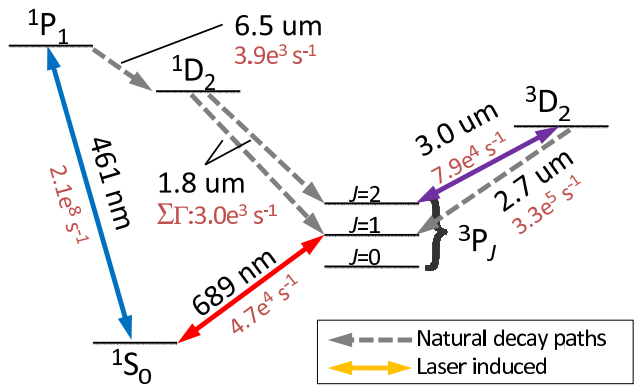}\\ %{node86.eps}\\
  \caption{Atomic Sr energy levels involved in the two-photon PAS experiments (color online).
  Decay rates (s$^{-1}$)
   and  excitation wavelengths are given for selected transitions.
  Laser light used for the experiment is indicated by solid lines.
  Atoms decaying to the $^3P_2$ level may be repumped by $3$\,$\mu$m light.
  %All frequencies
%are measured with respect to the atomic $^1S_0$-$^1P_1$ transition.
  }\label{levels}
\end{figure}

\subsection{Laser Cooling and Trapping}
%\subsubsection{}
To perform two-photon spectroscopy, we start with laser-cooled
atoms, and the initial cooling and trapping phases of the experiment
are similar to previously published descriptions
\cite{nsl03,nms05,mms05}.
%$^{88}Sr$ are
%initially trapped in a magneto-optical trap (MOT) operating on the
%461 nm $^1S_0$-$^1P_1$ transition.
%pure samples of each isotope
%from the same atomic beam due to the intrinsic isotope selectivity
%of a MOT.
Atoms are trapped in a magneto-optical trap (MOT) operating on the
461 nm $^1S_0$-$^1P_1$ transition (Fig.\ \ref{levels}) and cooled to
about 2\,mK. There is a decay channel from the $^1P_1$ state,
through the $^1D_2$ state, to the metastable $^3P_2$ level with a
branching ratio of $2 \times 10^{-5}$. To increase our sample
number, we repump  $^3P_2$ atoms by applying a 3\,$\mu$m  laser
resonant with the ${^3P_2}$-${^3D_2}$ transition that returns these
atoms to the ground state. The repumped sample of atoms contains
about $3.5 \times 10^8$ atoms.
%The 461 nm light used in this cooling stage
%and for the image beam is generated by frequency-doubling light from
%a titanium sapphire using KNb0$_3$ in a linear enhancement cavity
%\cite{bft97}.

After this initial MOT stage, the 461 nm light is extinguished and
the atom sample is transferred with more than 50\% efficiency to a
second MOT operating on the $^1S_0$-$^3P_1$ intercombination line
\cite{kii99}. The  sample is cooled to 3\,$\mu$K, producing
densities of $10^{12}$\,cm$^{-3}$. The 689 nm light is provided by a
master-slave diode laser system that is frequency-narrowed by
servo-locking it to a high-finesse optical cavity with the
Pound-Drever Hall method to produce a laser linewidth of $\sim
50$\,kHz. Long-term stability is maintained with a $^1S_0$-$^3P_1$
saturated-absorption cell.

\subsection{Crossed Optical Dipole Trap}

 To obtain high density and long sample lifetimes
 for improved two-photon PAS, atoms are transferred to an optical
 dipole trap (ODT) generated from a 21 W, 1064 nm, linearly-polarized,
 single-transverse-mode fiber laser.
 The trap is in a
crossed-beam configuration, derived from the first order deflection
of an acousto-optic modulator (AOM). The beam is focused on the
atoms with a minimum e$^{-2}$ intensity-radius of  $w$=75\,$\mu$m.
It is then  reflected back through the chamber to intersect the
first beam at 90 degrees and refocused to have the same waist at the
atoms. Both beams lie in a plane that is inclined $10.5$ degrees
from horizontal. The ODT trapping potential is calculated from
measured laser beam parameters and the polarizability of the $^1S_0$
state \cite{ykk08}. This allows us to determine the sample density
profile from the temperature and number of trapped atoms.
%The AOM allows fast turn
%on/off times  and power control of the dipole trap beams.

The maximum transfer efficiency observed from intercombination-line
MOT to ODT for an optimized overlap time of 80 ms between the two is
about 15\%. This is limited in large part by inelastic collisions
induced by 689 nm light.
%Up to $20 \times 10^6$  atoms are loaded into a single-beam
%ODT power of 5 W.
Atoms are initially loaded with a single-beam ODT power of 5\,W,
which creates a trap depth of about $U_{max}/k_B=25\,\mu$K. After
the 689 nm light is extinguished, the power is ramped to a final
value between 2.5 and 13\,W in 20 ms, yielding equilibrium
temperatures of between of 3 and 15\,$\mu$K. Up to $20 \times 10^6$
atoms  are loaded to yield peak densities on the order of
$10^{14}$\,cm$^{-3}$.
%The optimized transfer efficiency and stability of our atom sample
%also depends strongly on the optimized alignment of the FORT beams
%focused very near the center of the red MOT.
%Initially up to $17 \times 10^6$ ($8 \times 10^6$) are trapped in
%our deepest trap of xx uK (Do we always trap in the same trap. I
%don't get this...)
%and their equilibrium temperatures are about 7 uK
%(6 uK) . The sample size for our deepest trap is less than 200 um
%1/e size. We see rapid loss of our 86Sr samples, evident of a large
%3-body recombination rate. Typical lifetimes for the different
%isotopes are about 2 s (xx s) in the deepest trap, due to background
%gas collisions.  These parameters amount to a peak phase space
%density of about 0.06 in our deepest trap for either isotope.

The number of atoms and sample temperature  are determined with
time-of-flight absorption imaging using the $^1S_0$-$^1P_1$
transition. The lifetime of atoms in the ODT due to collisions with
background atoms is about $2\,\textrm{s}$.

%Trap oscillation frequencies measured by the  parametric resonance
%technique [PRA 57, R20 (1998)] agree with values calculated from the
%ODT laser power and measured beam waist,
%giving us confidence that we know the geometry of our trap.%confirm our ODT characterization. For a
%single-beam power of 6 W, we measure $\nu_{trap}=215$ Hz in the
%gravitational axis and 155 Hz in the perpendicular axes.

% Another important fact
%about our dipole trap is that the entrance angle of the trap beam
%with respect to horizontal is 10.5°, slightly complicating our trap
%potential, as gravity becomes a gradient.

\begin{figure}%%%%%Use Screen size centered on page when exporting from matlab!!!
 %%THis saves the sizes%
%print to pdfwriter from ppoint,open resulting pdf with ghostview, convert to
%eps with pswrite, max resoltuion
  %\includegraphics[width=3in,clip=true,trim=100 290 50 80 ]{energydiagramstraight.eps}\\
  %\includegraphics[width=3in,clip=true]{PASdrawing.eps}\\
  \includegraphics[width=4in,clip=true, trim=00 250 0 0, angle=0]{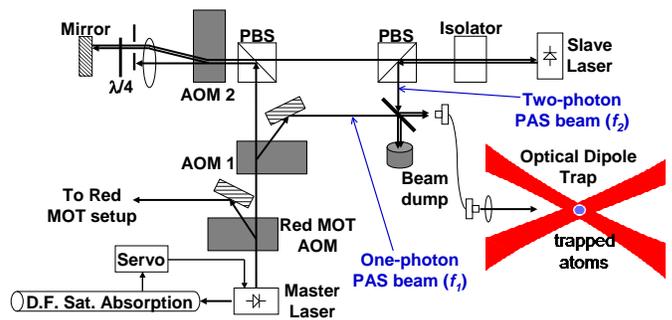}\\ %{node86.eps}\\
  \caption{Photoassociation lasers (color online). The master laser that provides light
  for the intercombination-line MOT is frequency-stabilized
  via saturated absorption
  spectroscopy to the atomic transition, and it
  also provides the photoassociation lasers.
 The one-photon PAS beam, with frequency $f_1$, is generated directly
  from the
  master with an AOM.
   The two-photon PAS beam, with frequency $f_2$, is
 formed by injection-locking a slave diode with a double-passed
 deflected beam from an AOM in a cat's eye configuration.
  %All frequencies
%are measured with respect to the atomic $^1S_0$-$^1P_1$ transition.
  }\label{Experimental Setup}
\end{figure}

\subsection{Photoassociation}

After the atoms have equilibrated in the ODT, the photoassociation
lasers are applied (Fig.\ \ref{PASDiagram}). Laser $f_1$ is near
resonance with a single-photon, free-bound transition to the red of
the $^1S_0$-$^3P_1$ atomic transition.  For some studies, this is
the only laser applied. For two-photon PAS, laser $f_2$ is near
resonance with a transition from the excited molecular bound state
to a ground molecular level.

%Laser $f_1$ is near resonant with a single-photon, free-bound
%transition 222.2\,MHz the red of the $^1S_0$-$^3P_1$ atomic
%transition. This is $J=1$ rotational state of the third least-bound
%vibrational level on the $0u$ molecular potential \cite{zbl06}.

%For $^{88}$Sr, $f_1$ is tuned near the -222.2\,MHz single-photon
%resonance
%In the case of 86Sr, we are currently searching for the
%single-photon PAS resonances, trying to attribute the signals we
%have measured to specific molecular potentials…

 Using acousto-optic modulators, all photoassociation
lasers are derived   from the master laser that provides the
intercombination-line MOT beams (Fig.\ \ref{Experimental Setup}).
Both $f_1$ and $f_2$ lasers are coupled into the same single-mode
optical fiber with the same linear polarization. More than 85\% of
the intensity of both output beams is linearly polarized in the
vertical direction, perpendicular to the ODT laser polarization.
These beams are focused to $w=200\,\mu$m
  at the location of the atom sample, which is substantially larger
than the atom cloud. The powers are monitored by a photodiode after
the fiber. Alignment of the $f_1$ and $f_2$ beams is facilitated by
co-propagating the beams with 461\,nm light aligned on the ODT atom
sample by absorption imaging using an independent CCD camera.
%Bringing $f_1$ on resonance with the
%single-photon PAS transition and aligning the beam to maximize the
%atom loss allows us to optimize the alignment of the fiber.
Depending on the specific measurement, one of the lasers is scanned
and the number of atoms remaining after the photoassociative
interaction time is recorded to obtain the loss spectrum.

%To trace out the two-photon PA spectrum at -135 MHz, f1 was held on
%resonance at -222.2 MHz while f2 frequency was stepped, and after
%75ms of two-photon photoassociation at a fixed frequency, the atoms
%are released from the FORT for about 10 ms and illuminated with a 80
%us 461 nm light pulse to count the number of atoms remaining.  To
%detect the Autler-Townes spectrum, f2 was held on resonance with the
%bound-bound transition and f1 was scanned near the free-bound
%transition.

 % Data from 15April2008: file 2253 for 88Sr (7.5V ODT), file 2731 for 86Sr (7.5V ODT)
 % Data from 15April2008: file 2253 for 88Sr (7.5V ODT), file 2723 for 86Sr (7.5V ODT)

\section{Theoretical Description of Photoassociative Loss
\label{Theoretical Description of Photoassociative Loss}}
%\subsection{}
%\subsubsection{}

Photoassociation is monitored by measuring the loss of ground-state
atoms from the ODT. This loss is described with a local equation for
the evolution of the atomic density
\begin{equation}\label{densitydecay}
\dot{n}=-2 Kn^2-\Gamma n,
\end{equation}
where the laser-frequency dependence of the collision event rate
constant, $K$, determines the spectrum of  the photoassociative
loss. The observed PAS spectrum is relatively simple because the
bosonic isotopes of strontium lack hyperfine structure. As shown in
Fig.\ \ref{PASDiagram}, ground state $^1S_0$ atoms collide on a
single $^1\Sigma^+_g$ potential. Four molecular potentials converge
to the $^1S_0$ + $^3P_1$ asymptote \cite{mjs01}, but only  states of
the $0_u$ and $1_u$ potentials are optically excited from the
$^1\Sigma^+_g$ potential \cite{zbl06}. At the low temperatures of
atoms in the ODT, only $s$-wave collisions occur so only $J=1$
intermediate levels and $J=0$ and 2 final states are populated.

%bound by 222.161\,MHz \cite{zbl06} is utilized in this work.

 Photoassociative loss
 %(assuming the level structure of Fig.\ 1 of \cite{bju96})
can be analyzed with the theory of Bohn and Julienne \cite{bju96},
which yields
\begin{equation}\label{Kintegral}
   K=\frac{1}{h\,Q_{T}} \int \vert S(E_g,f_1,f_2,...)\vert^2
   \,e^{-E_g/k_{B}T} \; dE_g
\end{equation}
where the partition function is $Q_{T}=\left({2\pi k_{B}T \mu \over
h^2}\right) ^{3/2}$ for reduced mass $\mu$. In spite of the low
temperature, thermal averaging over the collision energy $E_g$ is
necessary because of the narrow linewidth of the transition. $\vert
S\vert^2$ is the scattering probability for loss and its structure
depends upon the loss process that is dominant.

  In two-photon
spectroscopy of alkali-metal systems,
 the dominant photoassociative loss process is a collision on the open
channel of two ground state atoms (g) with total energy $E_g$
leading to loss-producing decay from the excited state $b_1$  with
rate $\gamma_1$. (See Fig.\ \ref{PASDiagram}.) However, in the
experiment reported here, $b_1$ is metastable, and there is a
concern that loss from the ground molecular state $b_2$ may also be
important. The complete vanishing of the photoassociative loss when
the lasers are on two-photon resonance from $g\rightarrow b_2$
(Secs.\ \ref{section Autler Townes} and \ref{sectionTwo-Photon
Suppression}), however, implies that decay from $b_2$ is negligible
for our conditions and loss from $b_1$ dominates. We can express the
scattering probability as
\begin{eqnarray}\label{equationSprob}
  \vert S_{1g}\vert^2 =   \hspace{2.5in}                          &&\\
  {\left(\delta_1-\delta_2\right)
^2{\gamma}_1{\gamma}_s/(2\pi)^2 \over \left\{ \left[
\delta_1-\frac{\delta_2}{2}\right] ^2-\frac{1}{4}\left[
\delta_{2}^{2}+\frac{\Omega_{12}^{2}}{(2\pi)^2}\right]
\right\}^2+\left( \frac{{\gamma}_1+{\gamma}_s}{4\pi}\right)^2\left(
\delta_1-\delta_2\right) ^2}. &&\nonumber
\end{eqnarray}
%\begin{equation}\label{equationSprob}
%\vert S_{1g}\vert^2 = {\left( \frac{E}{h}+\delta_1-\delta_2\right)
%^2{\gamma}_1{\gamma}_s/(2\pi)^2 \over \left[ \left(
%\frac{E}{h}+\delta_1-\frac{\delta_2}{2}\right) ^2-\frac{1}{4}\left(
%\delta_{2}^{2}+\frac{\Omega_{12}^{2}}{(2\pi)^2}\right)
%\right]^2+\left( \frac{{\gamma}_1+{\gamma}_s}{4\pi}\right)^2\left(
%\frac{E}{h}+\delta_1-\delta_2\right) ^2}.
%\end{equation}
${\gamma}_{1}=2\gamma_{atomic}$, where $\gamma_{atomic}$ is the
decay rate of the atomic $^3P_1$ level, and ${\gamma}_{s}(E_g)$ is
the stimulated width of $b_1$ due to laser-coupling to $g$,
\begin{equation}\label{equationstimulatedwidth}
{\gamma}_{s}(E_g)=\frac{2\pi V^{2}\vert\langle b_{1}\vert E_g\rangle
\vert ^2}{\hbar},
\end{equation}
where  we represent $g$ as the energy-normalized colliding state
$\vert E_g\rangle$, leading to the  the Franck-Condon factor for the
free-bound transition, $\vert\langle b_{1}\vert E_g\rangle\vert ^2$,
and
%$V=\left( \frac{2\pi
%I_{1}}{c}\right)^{1/2}d$ in CGS,
$V=d \left( \frac{I_{1}}{2\epsilon_{0}c}\right)^{1/2}$ for
free-bound laser intensity $I_1$ and molecular dipole matrix element
$d$. Note that our $\Omega_{12}$ is the splitting of the
Autler-Townes doublet (Sec.\ \ref{section Autler Townes}), which
differs from the Bohn-Julienne definition of the molecular Rabi
coupling \cite{bju96}.

%
%$
%\Omega_{12}^2=\Omega_{C-T,mol}^{2}= f b_{12}^{2}\gamma_{atom}^{2}
%{\frac{2I}{I_{sat,atom}}} =4\Omega^2_{B-J}.
%$
%$f=\vert\langle b_{1}\vert b_{2}\rangle \vert ^2$ is the
%Frank-Condon factor between the bound-bound states. $b_{12}$
%accounts for the change in dipole moment from atom to molecule due
%to symmetry of wave function and projection on rotating molecular
%axis.
%
%\begin{equation} \hbar\Omega_{mol}=\hbar b_{12} \Omega_{atom} \qquad
%\qquad 0<b_{12}<\frac{2}{\sqrt{3}}
%\end{equation}
%$\frac{2}{\sqrt{3}}$ is usually good.  WE NEED TO CHECK.

The thermal energy is much greater than the zero-point energy for
trap motion, $T>>h\nu_{trap}/k_B$, so confinement effects are
negligible \cite{zbl06}. We also neglect Doppler shift and photon
recoil \cite{ctk04}, which is reasonable since $T>T_R$, where the
recoil temperature for $\lambda=689$\,nm photons is
$T_R=\textit{h}^2/(k_B\lambda^2 m)=460$ nK. We assume the decay
products leave the trap, which is a good approximation for the
intermediate levels we use, although it is not for the least bound
$0_u$ excited molecular
state  \cite{zbl06,jul08}. %in $^{88}$Sr%

The energy integral for $K$ (Eq.\ \ref{Kintegral}) is not analytic
and must be evaluated numerically.  The situation is further
complicated by the ODT, which is not at the magic wavelength
 for one-photon photoassociation (914\,nm \cite{zbl06}).
 The AC Stark shift of the weakly bound ground molecular level
($b_2$) is approximately equal to the shift of the incoming channel
of 2-free atoms ($g$) \cite{rbm04,rfk87}. In other words, the
polarizability of the ground molecule is about twice that of a
single atom. But the ground and excited molecular levels do not
experience the same  shift. For spectroscopy, we can thus treat the
ODT Stark shifts as a position-dependent shift of the intermediate
state and define the laser detunings
\begin{eqnarray}
\label{equationlightshift}
   \delta_{1} &=& f_{1}-(E_{b_1}- E_g)/h-\chi I_{ODT}(\vec{r})\nonumber \\
  \delta_{2} &=& f_{2}-(E_{b_1}-E_{b_2})/{h} -\chi I_{ODT}(\vec{r}),
\end{eqnarray}
where $I_{ODT}(\vec{r})$ is the intensity profile of the optical
dipole trap and $\chi$ can be related to the differences in
polarizabilites for $^1S_0$ and $^3P_1$ atoms for 1.06\,$\mu$m laser
light.

This implies that $\vert S\vert^2$ and thus $K$ are functions of
position, which must be addressed when Eq.\ \ref{densitydecay} is
integrated over the trap volume to calculate the time evolution of
the number of trapped atoms
% Now $\vert S\vert^2$ is a function of $\delta_{1}$, $\delta_{2}$,
%$E$, $\vec{r}$ (and other things).
%\begin{equation}
%\vert S(\delta_{1},\delta_{2},E,\vec{r})\vert^2 \Rightarrow
%K(\delta_{1},\delta_{2},E,\vec{r})
%\begin{equation}\fbox{ $\displaystyle \delta_{1} \rightarrow \delta_{1}-\chi
%I(\vec{r})\qquad \delta_{2}  \rightarrow  \delta_{2}-\chi
%I(\vec{r}). $ }\end{equation} Now $\vert S\vert^2$ is a function of
%$\delta_{1}$, $\delta_{2}$, $E$, $\vec{r}$ (and other things).
%\begin{equation}
%\vert S(\delta_{1},\delta_{2},E,\vec{r})\vert^2 \Rightarrow
%K(\delta_{1},\delta_{2},E,\vec{r})
%\end{equation}
\begin{equation}\label{number}
   N(t)={N_0 \rm{e}^{-\Gamma t} \over 1+
   {2 N_0 K_{eff}V_2\over \Gamma V_1^2}(1-\rm{e}^{-\Gamma t})},
\end{equation}
where  $N_0$  is the number  at the beginning of the PAS interaction
time. The one-body loss rate, $\Gamma$, is due to background
collisions and off-resonant scattering from the PAS lasers. The
effective volumes are defined by
\begin{equation}\label{eq:effectivevolumes}
V_{q}=\int_{\mathrm{V}} d^3r \, e^{-\frac{qU(\vec{r})}{k_{B}T}},
\end{equation}
where $U(r)$ is the trap potential, and
%\begin{equation}
% K_{eff}=\frac{1}{V_{2}}\int_{\mathrm{V}}
%\mathrm{d}^3r \,
%          e^{-\frac{2U(\vec{r})}{k_{B}T}}K(\vec{r})
%\end{equation}
%\begin{equation}\label{equationKeffective}
% K_{eff}=\frac{1}{V_{2}}\int_{\mathrm{V}}
%\mathrm{d}^3r \,
%          e^{-\frac{2U(\vec{r})}{k_{B}T}}\frac{k_{B}\,T}{h\,Q_{T}}
%\int_{0}^{U_{max}-U(r)}\frac{\mathrm{d}E} {k_{B}\,T} \vert
%S_{1g}\vert^2
%   \,e^{-E/k_{B}T}.
%\end{equation}
\begin{eqnarray}\label{equationKeffective}
  K_{eff}&=& \frac{1}{V_{2}}\int_{\mathrm{V}}
d^3r \,
          e^{-\frac{2U(\vec{r})}{k_{B}T}} \nonumber \\
         &&\times \frac{1}{h\,Q_{T}}
\int_{0}^{U_{max}-U(r)}dE_g \vert S_{1g}\vert^2
   \,e^{-E_g/k_{B}T}.
\end{eqnarray}
%\begin{eqnarray}\label{equationKeffective}
%% \nonumber to remove numbering (before each equation)
%  K_{eff}=  \hspace{2.8in}                          &&\\
%  \frac{1}{V_{2}}\int_{\mathrm{V}}
%\mathrm{d}^3r \,
%          e^{-\frac{2U(\vec{r})}{k_{B}T}}\frac{k_{B}\,T}{h\,Q_{T}}
%\int_{0}^{U_{max}-U(r)}\frac{\mathrm{d}E} {k_{B}\,T} \vert
%S_{1g}\vert^2
%   \,e^{-E/k_{B}T}. &&\nonumber
%\end{eqnarray}
The  kinetic energy integral is truncated by the local trap depth,
$U_{max}-U(r)$. The spatial integrals in Eqs.\
\ref{eq:effectivevolumes} and \ref{equationKeffective} extend over
the trap volume V in which $U(r)<U_{max}$. Atom temperatures vary by
no more than 25\% during the interaction time, so assuming a
constant sample temperature is reasonable.

%MAIN PARAMETERS:  We control \begin{center} $\begin{array}{l}
%\delta_{1} (through f_{1}) \\
%\delta_{2} (through f_{2}) \\
%I_{1} \\ I_{2}
%\end{array}$\end{center}
%and to some extend the trap, $U(\vec{r})$, and the temperature.  We
%hope to determine (or take as given)
%
%\begin{tabular}{cl}
%  $E_{b_{1}}=E_{^{3}\!P_{1}}-222.16$ MHz & given binding energy of excited molecular level \\
%  $\chi \left[ \frac{\mathrm{Hz}}{\mathrm{W}}\right]$ & shift of 1-photon PAS per ODT intensity \\
%  $f=\vert\langle b_{1}\vert b_{2}\rangle \vert ^2$ & bound-bound Frank-Condon factor\\
%  $B$ from $\tilde \gamma_{2}=BI_{1}\sqrt{E}$ & is related to the
%  free-bound Frank-Condon factor
%\end{tabular}
%
%KEY: $E_{b_{2}}$ is the \underline{binding energy of ground state
%molecules}.

The spectrum is sensitive to many atomic and molecular parameters,
and  multiple types of spectra  can be used to determine them. The
ultimate goal is an accurate determination of $E_{b_2}$ because the
molecular binding energy  determines the $s$-wave scattering length
and the underlying potential with high accuracy.

\section{One-Photon Photoassociation\label{One-Photon Photoassociation}}
%\subsection{}
%\subsubsection{}
\label{sectiononephoton}

\subsection{One-Photon PAS Spectrum}

One photon PAS allows us to determine $\chi$, the relative light
shift of states on the ground and excited molecular potentials (Eq.\
\ref{equationlightshift}), and the stimulated width
${\gamma}_{s}(E_g)$ (Eq.\ \ref{equationstimulatedwidth}).

When $I_{2}=0$ and $\Omega_{12}=0$ in Eq.\ \ref{equationSprob}, we
recover the one-photon PAS scattering probability for loss through
decay of $b_1$
\begin{equation}\label{equationSprob1}
\vert S_{1g}\vert^2 = {{\gamma}_{1}{\gamma}_{s}/{(2\pi)^2}\over
\delta_1^2 +\frac{1}{(2\pi)^2}\left(
\frac{{\gamma}_1+{\gamma}_s}{2}\right)^2}.
\end{equation}
%\subsection{Transition Utilized}
The state $b_1$  is equal to the $J=1$ rotational state of the
third-least bound vibrational level of the $0_u$ potential, with
energy $E_{b_1}=E_{^3P_1}- h \times 222.161(35) $\,MHz \cite{zbl06}.
The Condon point for this excitation, where
$V_{0_u}(R_c)-V_{^1S_0}(R_c)=E_{b_1}$, occurs at $R_c=75$\,a$_0$,
which is very near the node in the ground
state wave function \cite{mms05,ykt06}. %This makes the transition strength
%sensitive to the details of the ground state potential \cite{zbl06}
%and relatively weak. \textbf{(CHECK)}

\begin{figure}%%%%%Use Screen size centered on page when exporting from matlab!!!
 %%THis saves the sizes%
%print to pdfwriter from ppoint,open resulting pdf with ghostview, convert to
%eps with pswrite, max resoltuion
  %\includegraphics[width=3in,clip=true,trim=100 290 50 80 ]{energydiagramstraight.eps}\\
  %\includegraphics[width=3in,clip=true]{PASdrawing.eps}\\
  \includegraphics[width=3.5in,clip=true, trim=0 0 0 0, angle=0]{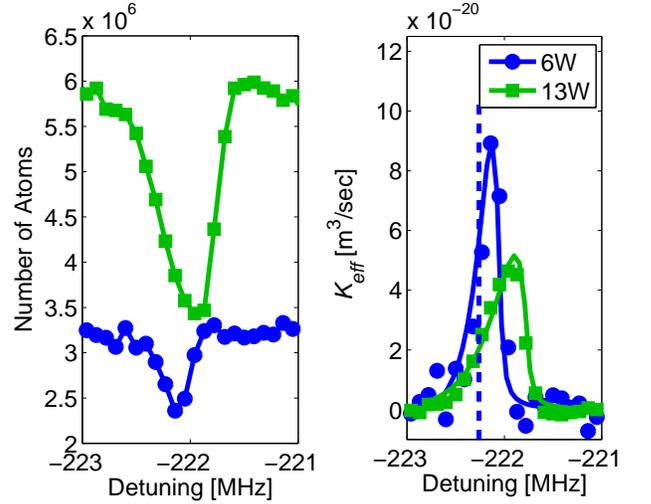}\\ %{node86.eps}\\
  \caption{Left: Atom number versus free-bound laser detuning from the one-photon
  $^1S_0$-$^3P_1$ atomic transition (color online).
  Spectra shown here are for 6 W and 13 W ODT single-beam powers with sample
  temperatures of $6\,\mu$K and $13\,\mu$K, respectively.
  Right: Collision-event rate constant $K_{eff}$ derived from the atom loss.
  The ODT at 1064 nm causes an AC Stark shift of the excited molecular state
  compared to the ground state, which shifts and broadens  the line.
  The solid lines are
  fits using Eqs.\ \ref{equationKeffective} and
  \ref{equationSprob1}.  A peak shift of 480\,kHz is measured for a
  single-beam power of 13 W.
The dashed line marks the position of our measured unperturbed
resonance frequency at -222.25(15) MHz,
  which is in reasonable agreement with a previous measurement of -222.161(35) MHz \cite{zbl06}.
  }\label{OnePhotonPAS}
\end{figure}

Fitting data of atom number after a given interaction time, $N(t)$,
to Eq.\ \ref{number} yields the collision-event rate constant
$K_{eff}$.  Figure \ref{OnePhotonPAS} shows typical spectra for this
transition.

 \begin{figure}%%%%%Use Screen size centered on page when exporting from matlab!!!
 %%THis saves the sizes%
%print to pdfwriter from ppoint,open resulting pdf with ghostview, convert to
%eps with pswrite, max resoltuion
  %\includegraphics[width=3in,clip=true,trim=100 290 50 80 ]{energydiagramstraight.eps}\\
  %\includegraphics[width=3in,clip=true]{PASdrawing.eps}\\
  \includegraphics[width=3in,clip=true, trim=00 0 0 0, angle=0]{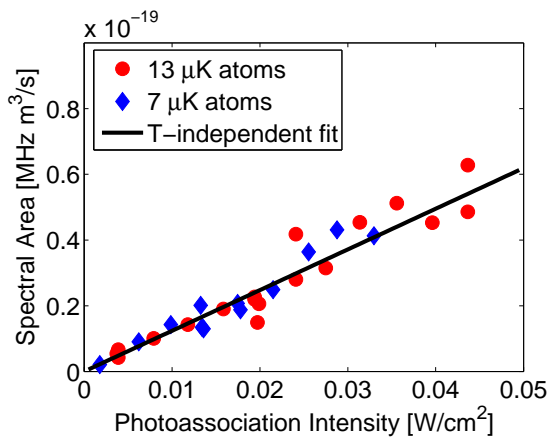}\\ %{node86.eps}\\
  \caption{Area under one-photon PAS spectra versus free-bound
  laser intensity $I_1$ (color online).  The area can be related to molecular and
  experimental parameters to determine the stimulated linewidth ${\gamma}_{s}(E_g)$ of the
  PAS transition due to $I_1$.
  For low free-bound laser intensities the area is independent of
  temperature and linearly dependent on $I_1$.
  %We need a caption and the fonts, lines, and data symbols
  %need to be made easier to see. What information should go in the
  %legend? What should the axes labels be? Resolution needs to be
  %better? Can we put this in terms of K instead of beta?
  %All frequencies
%are measured with respect to the atomic $^1S_0$-$^1P_1$ transition.
  }\label{OnePhotonPASArea}
\end{figure}

\subsection{Determining the Stimulated Width }

The area under the one-photon PAS line (Fig.\ \ref{OnePhotonPAS})
can be related to molecular and experimental parameters through
\begin{eqnarray}
  A &=& \int df K_{eff} \nonumber \\
   &=& \frac{1}{h\,Q_{T}}
\int dE_g
   \,e^{-E_g/k_{B}T}\frac{\gamma_{s}(E_g)\gamma_1}{\gamma_{s}(E_g)+\gamma_1}.
\end{eqnarray}
Here, we have neglected truncation of the energy integral, which is
a small correction. The Wigner threshold law implies
${\gamma}_{s}(E_g) \propto I_{1}\sqrt{E_g}$. So for low laser
intensity ($\gamma_{s}(E_g)\ll \gamma_1$), the expression $A \approx
\frac{k_{B}\,T}{h\,Q_{T}} \left\langle\gamma_{s}(E_g)\right\rangle$
is independent of temperature and linearly dependent on intensity
(Fig.\ \ref{OnePhotonPASArea}), where
$\left\langle\cdots\right\rangle$ refers to a thermal average. The
PAS saturation intensity for this transition ($I_{sat,PAS}\propto
1/T^{1/2}$) defined as the intensity for which $\left\langle
\gamma_{s}(E_g)\right\rangle=\gamma_1$, is $8$\,W/cm$^2$ for
$T=13\,\mu$K. Expressing this in terms of an optical length for the
transition $\ell_{opt}=\frac{\gamma_{s}(E_g)}{2 k_r\gamma_1}$, where
$k_r=\sqrt{2 \mu E_g}/\hbar$, yields $\ell_{opt}=28$\,$a_0$ for
$I_1=1$\,W/cm$^2$ for this transition. %(Need uncertainties) (For
%B=$1.0\times 10^{14}$).

 %determines
%\underline{B}, or $\tilde{\gamma}_{s}$. one can show that for
%$\gamma_{2}<<\gamma_{1}$ ($I_{1}<<I_{sat}$) [$A$ is in units of
%$\frac{\mathrm{m}^3}{\mathrm{s}}\mathrm{Hz}$],
%
%\begin{equation}
%A=2\frac{k_{B}\,T}{h\,Q_{T}}\frac{1}{2} \langle
%\tilde{\gamma}_{s}(E) \rangle
%\end{equation}
%where (neglecting trap limit),
%\begin{equation}
%\langle \tilde{\gamma}_{s}(E) \rangle=\int \frac{\mathrm{d}E}
%{k_{B}\,T} \,\tilde{\gamma}_{s}(E)
%   \,e^{-E/k_{B}T}.
%\end{equation}
%
%\begin{equation}
%A=\frac{(k_{B}\,T)^{3/2}}{h\,Q_{T}} B\, I \frac{\sqrt{\pi}}{2}.
%\end{equation}
%Saturation intensity is when $\langle \tilde{\gamma}_{s}(E) \rangle
%= \tilde{\gamma}_{1}$,
%\begin{equation}
%\Rightarrow I_{sat}={4 \gamma_{atom} \over B \sqrt{\pi k_{B}\,T}}
%\end{equation}

\subsection{Modelling the Spectra and Determining the Relative AC Stark Shift}
\label{subseclightshift}
 Numerical integration of Eq.\ \ref{equationKeffective} to find $K_{eff}$
  using Eq.\ \ref{equationSprob1} for the scattering probability allows
 us to model the one-photon PAS spectra, and  the relative light
shift parameter (Eq.\ \ref{equationlightshift}) can be varied to fit
the data (Fig.\ \ref{OnePhotonPAS}). We find $\chi=160 \pm 30$\,kHz/
(100 kW/cm$^2$), in good agreement with \cite{ykk08}, which yields a
peak shift of 480\,kHz for our deepest trap. The line shifts to the
blue with more ODT laser intensity, showing the polarizability of
$^3P_1$ atoms is less than the polarizability of $^1S_0$ atoms.

At higher temperatures, the lineshapes in Fig.\ \ref{OnePhotonPAS}
possess red tails, which result from the convolution of the
Lorentzian spectrum with the Maxwell-Boltzmann distribution of
collision energies \cite{zbl06,ctk04}.
%Equation \ref{Kintegral} neglects Doppler
%broadening, which is a reasonable approximation at for our sample
%temperatures \cite{ctk04}.

\section{Two-Photon Autler Townes Spectrum}
\label{section Autler Townes}
%\subsection{}
%\subsubsection{}

If PAS spectra are recorded in the same fashion as in Sec.\
\ref{sectiononephoton}, except the bound-bound laser is added near
resonance ($\delta_2\approx 0$) with a large intensity $I_2$, the
loss spectrum is modified due to the coupling between $b_1$ and
$b_2$. This forms a $\Lambda$ system and the line is split into an
Autler-Townes doublet, with splitting  given approximately by
$\Omega_{12}/2\pi$,
%$\Omega_{12}/2\pi \propto
%\sqrt{I_2}\left|\langle b_1|b_2\right\rangle|$,
where
%\begin{equation}\label{equationmolecularrabifreq}
%    \Omega_{12}=\alpha \vert\langle b_{1}\vert
%b_{2}\rangle \vert \gamma_{atom} \sqrt{{\frac{2I}{I_{sat,atom}}}}.
%\end{equation}
\begin{equation}\label{equationmolecularrabifreq}
    \Omega_{12}=\alpha \vert\langle b_{1}\vert
b_{2}\rangle \vert \gamma_{1} \sqrt{{\frac{I}{4I_{sat,atom}}}}.
\end{equation}
The overlap integral is related to the Franck-Condon factor, $F$,
through $F=\vert\langle b_{1}\vert b_{2}\rangle \vert ^2$. The
saturation intensity for the atomic $^1S_0-^3P_1$ transition is
$I_{sat,atom}\equiv \pi h c
\gamma_{atomic}/(3\lambda^3)=3\,\mu$W/cm$^2$. The rotational line
strength factor, $\alpha$, accounts for the change in dipole moment
from atom to molecule due to symmetry of wave function and
projection on a rotating molecular axis
\cite{nju06,mjs01}. %\textbf{$0<\alpha<\frac{2}{\sqrt{3}}$ and
%$\frac{2}{\sqrt{3}}$ is usually good.  WE NEED TO CHECK. }

 \begin{figure}%%%%%Use Screen size centered on page when exporting from matlab!!!
 %%THis saves the sizes%
%print to pdfwriter from ppoint,open resulting pdf with ghostview, convert to
%eps with pswrite, max resoltuion
  %\includegraphics[width=3in,clip=true,trim=100 290 50 80 ]{energydiagramstraight.eps}\\
  %\includegraphics[width=3in,clip=true]{PASdrawing.eps}\\
  \includegraphics[width=3.5in,clip=true, trim=50 60 0 0, angle=0]{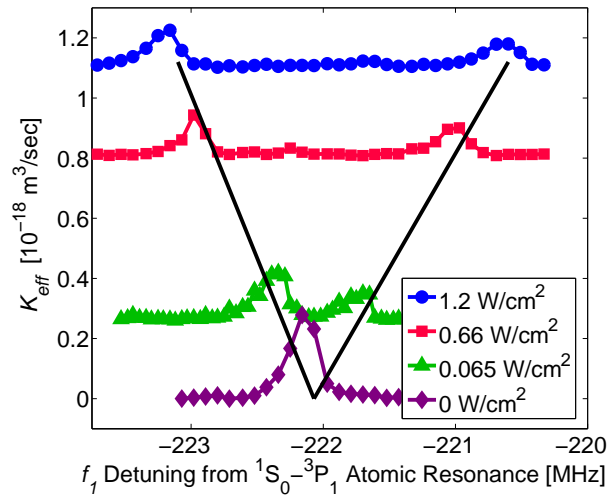}\\ %{node86.eps}\\
  \caption{Collision-event rate constant $K_{eff}$ versus free-bound
  laser detuning from $^1S_0$-$^3P_1$ atomic resonance for different bound-bound laser intensities (color online).  These
  Autler-Townes doublets are measured with the bound-bound laser
  frequency fixed such that $\delta_2\approx 0$ while scanning $f_1$.
  The splitting of the spectra is given by the Rabi frequency $\Omega_{12}/2
  \pi$ and varies as $\sqrt{I_2}$ (Shown by the lines; spectra offset is
  proportional to $\sqrt{I_2}$), where the bound-bound laser intensity $I_2$ is
  indicated in the legend.  The
  asymmetry in the lineshapes arises from the bound-bound laser
  frequency being slightly off resonance from the bound-bound
  transition. The free-bound intensity $I_1$ is constant for all four spectra at
   0.05 W/cm$^2$. The sample temperature is  $8 \mu$K.
  %Better fonts, bigger points, etc.. What should go in the
  %caption? We can use the offset spectra showing the Rabi frequency
  %scaling.
  %All frequencies
%are measured with respect to the atomic $^1S_0$-$^1P_1$ transition.
  }\label{AutlerTownes}
\end{figure}

Figure \ref{AutlerTownes} shows several Autler-Townes spectra for
various intensities $I_2$ for a ground molecular state  ($b_2$)
equal to the  $J=0$ rotational state of the least-bound ground
vibrational level. This is the $v=62$ level counting from the bottom
of the well. We find that $\Omega_{12}/2\pi=1$ MHz for an intensity
of $.35$\,W/cm$^2$, which yields  $F=0.28\pm0.06$ for
$\alpha=\sqrt{2/3}$ \cite{mjs01}. %The fits to the spectra are the
%numerical evaluations of $K_{eff}$ (Eq.\ \ref{equationKeffective})
%including thermal and trap effects, with parameters as found in
%Sec.\ \ref{sectiononephoton}.
The  energy of the ground molecular state, $E_{b_2}$, can be found
from fits of data in Fig.\ \ref{AutlerTownes}, but it is more
accurately found by varying $\delta_2$ with $\delta_1\approx 0$, as
we will discuss in Sec.\ \ref{sectionTwo-Photon Suppression}.

The asymmetry in the line strengths in each doublet in Fig.\
\ref{AutlerTownes} arises from the coupling-laser frequency
 being slightly off resonance from the Stark-shifted
bound-bound transition ($\delta_2\neq0$). But $\delta_2$ is small
and the scaling with intensity shows that the Autler-Townes
splitting varies as $\sqrt{I_2}$ as expected.

%The origin of the Autler-Townes doublet can easily be understood in
%the dressed state picture of the atom. The bound-bound laser couples
%the third least-bound level of the $^1S_0$-$^3P_1$ molecular
%potential to the least bound level of the $^1S_0$-$^1S_0$ molecular
%potential and its intensity determines the dressed states'
%separation by the generalized Rabi frequency $\Omega_{12}/2 \pi$ of
%this transition. The increase of number loss (or in other words, the
%increase of the two-body loss rate in Fig.\ \ref{AutlerTownes})
%comes about when the weaker free-bound laser is scanned and becomes
%on resonance with one or the other of the two dressed states,
%causing a photoassociation transition.

These spectra show the potential of the system for quantum optics
and ultracold molecule formation. The condition of no PAS loss when
both lasers are on resonance has also been called a dark resonance
\cite{mpk06}, or an atom-molecule dark state \cite{wtt05} because
the system has been put in an atom-molecule superposition state with
vanishing excitation rate to $b_1$. Such a state has also been
proposed as a vehicle for creating large numbers of ground state
molecules using STIRAP \cite{mkj00}. The level of suppression near
$\delta_1=0$ shows the coherence of this superposition state in
these experiments. It is noteworthy that the doublet is split by
many linewidths even with moderate coupling-laser intensity because
of the intrinsically narrow spectrum of intercombination-line PAS.

\section{Two-Photon Suppression of Photoassociation
\label{sectionTwo-Photon Suppression}}
%\subsection{}
%\subsubsection{}

 \begin{figure}%%%%%Use Screen size centered on page when exporting from matlab!!!
 %%THis saves the sizes%
%print to pdfwriter from ppoint,open resulting pdf with ghostview, convert to
%eps with pswrite, max resoltuion
  %\includegraphics[width=3in,clip=true,trim=100 290 50 80 ]{energydiagramstraight.eps}\\
  %\includegraphics[width=3in,clip=true]{PASdrawing.eps}\\
  \includegraphics[width=3in,clip=true, trim=00 0 0 0, angle=0]{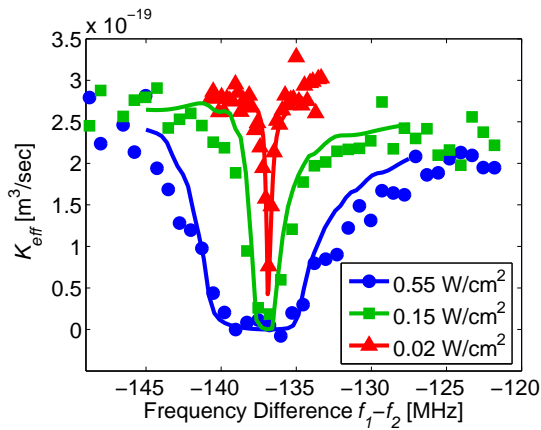}\\ %{node86.eps}\\
  \caption{Collision-event rate constant $K_{eff}$ versus frequency
  difference between free-bound and bound-bound
  lasers for spectroscopy of the $J=0$, $v=62$  level  of the
  X$^1\Sigma_g^+$ potential (color online).
  The free-bound laser frequency is fixed close to the
  one-photon PAS resonance and the intensity is $I_1=0.05$\,W/cm$^2$.
  The bound-bound laser frequency is
  scanned, and its intensity is indicated in the legend.
    On two-photon resonance, PAS loss is suppressed due
  to quantum interference.  The solid lines are
  fits using Eqs.\ \ref{equationKeffective} and
  \ref{equationSprob}, which yield $E_{b2}/h=-136.7(2)$\,MHz.
   The atom
temperature is  $8\, \mu$K.
  %Fonts, axes, legend.....pointsize is good.
  %All frequencies
%are measured with respect to the atomic $^1S_0$-$^1P_1$ transition.
  }\label{PASSuppression}
\end{figure}

For determining the binding energy of  molecular levels of the
ground state potential, we hold the frequency of the free-bound
laser fixed close to the one-photon resonance, $ \delta_1\simeq0$,
and scan $\delta_2$. When $\delta_2-\delta_1=0$, the system is in
two-photon resonance from state $g$ to $b_2$, and one-photon
photoassociative loss is suppressed due to quantum interference. At
this point, $f_1-f_2=(E_{b2}-E_g)/h$, so the spectrum allows
accurate determination of  $E_{b_2}$. An average over $E_g$ is
necessary in order to properly account for thermal shifts of the
resonance.

\begin{figure}%%%%%Use Screen size centered on page when exporting from matlab!!!
 %%THis saves the sizes%
%print to pdfwriter from ppoint,open resulting pdf with ghostview, convert to
%eps with pswrite, max resoltuion
  %\includegraphics[width=3in,clip=true,trim=100 290 50 80 ]{energydiagramstraight.eps}\\
  %\includegraphics[width=3in,clip=true]{PASdrawing.eps}\\
  \includegraphics[width=3in,clip=true, trim=0 0 0 0, angle=0]{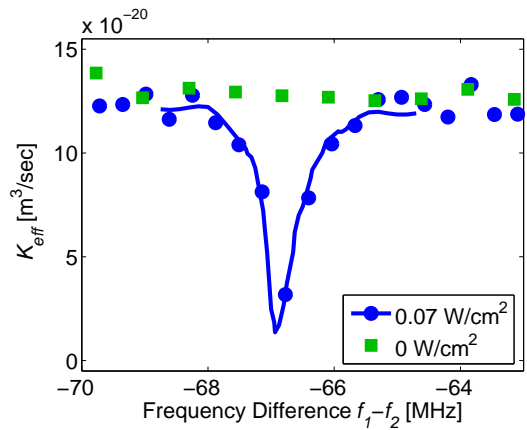}\\ %{node86.eps}\\
  \caption{PAS suppression spectra for the $J=2$, $v=62$  level  of the
  ground molecular potential, as described in Fig. \ref{PASSuppression} (color online).
    The sample temperature is $9\,\mu$K. The free-bound laser  intensity is $I_1=0.04$\,W/cm$^2$, and the
    bound-bound intensity is indicated in the legend.
  The solid line is a
  fit using Eqs.\ \ref{equationKeffective} and
  \ref{equationSprob}, which yields $E_{b2}/h= -66.6(2)$\,MHz.
  %All frequencies
%are measured with respect to the atomic $^1S_0$-$^1P_1$ transition.
  }\label{suppresionJ2}
\end{figure}

Figure \ref{PASSuppression} shows a series of spectra taken at
various bound-bound intensities for $b_2$ equal to  the  $J=0$,
$v=62$ state. Detuning of the free-bound laser frequency ($f_1$)
from the free-bound resonance, which depends on the ODT light shift
($\chi$) and collision energy ($E_g$), causes slight asymmetry in
the lines and broadening, but this shape is reproduced with our
model for $K_{eff}$ (Eq.\ \ref{equationKeffective}) using parameters
independently determined in previous sections.  Since the initial
and final states experience roughly equal light shifts due to the
trapping laser, the ODT AC Stark shifts do not shift the resonance.
No significant shift of the binding energy with laser power was
observed, and we place an upper limit of 100\,kHz for our highest
intensity, $I_2=0.55$\,W/cm$^2$. We have also measured the binding
energy of the $J=2$, $v=62$ state (Table
\ref{tablebindingenergies}), and a typical spectrum is shown in
Fig.\ \ref{suppresionJ2}.

%

%\begin{tabular}{|Isotope|v|J|E_b (MHz)|E_b (MHz)|Difference|}
\begin{table}
\begin{tabular}{|l|c|c|c|c|c|}
 \hline
 Isotope & $v$ & $J$ & Exp. & Theory & Diff. \\
   \hline
 88 & 62 & 0 & -136.7(2) & -136.7 & 0.0 \\
88 & 62 & 2 & -66.6(2) & -66.5 & -0.1 \\
  \hline
  \end{tabular}
\caption{Observed ground molecular levels and
  experimental and theoretical level energies in frequency units ($E_{b2}/h$).}
  \label{tablebindingenergies}
\end{table}

%\begin{table}
%\begin{tabular}{|l|c|c|c|c|c|}
% \hline
%  % after \\: \hline or \cline{col1-col2} \cline{col3-col4} ...
% isotopes & 2-phot. & Fourier-& 1-phot. & 1-phot. & thermal-  \\
%         & PAS &   transform & PAS        & PAS &  ization  \\
%  & this study & \cite{skt08}& \cite{mms05} & \cite{ykt06} & \cite{fdp06} \\
%    & $a$ & $a$& $a$ & $a$ & $|a|$ \\
%
%   \hline
% 88-88 & -1.2(.2) & $0(5)$ &$10^{+3}_{-11}$ & $-40^{+40}_{-100}$ & $21^{+3}_{-4}$\\
% \hline
%87-87 & 96() & $97(2)$     &-& - & - \\
%\hline
%86-86 & 833() & $1050(380)$ &$1000^{+1300}_{-400}$ & - & $430^{+80}_{-90}$ \\
% \hline
%84-84 & 123() &  $124(3)$    &-& - & - \\
% \hline
%88-87 & 55() & $56(2)$     &-& - & - \\
%\hline
%88-86 & 98() & 99(2)& - & - & $110^{+10}_{-20}$  \\
% \hline
%88-84 & 1845() & $>1170$ or    &-& - & - \\
%      &        & $<-1900$     & &   &   \\
%\hline
%87-86 & 163() &  165.5(5.5)    &-& - & - \\
%\hline
%87-84 & -56() &  -55(10)    &-& - & - \\
%\hline
%86-84 & 32() &  $33(3)$    &-& - & - \\
%  \hline
%  \end{tabular}
%\caption{Published values of scattering lengths given in units of
%$a_0=.053$\,nm}
%  \label{tablescatteringlengths}
%\end{table}
\section{Determination of  the Scattering Length\label{Determination of the Scattering Length}}
%\subsection{}
%\subsubsection{}

Binding energy measurements can be used to accurately determine the
$s$-wave scattering length for the X$^1\Sigma_g^+$ ground electronic
molecular potential and to obtain information on the van der Waals
coefficients, $C_n$. For our analysis, the inner part (internuclear
spacing $R<20.8$\,$a_0$) of the potential is described with a
recently published energy curve \cite{skt08} derived from the
Fourier-transform spectrum of Sr$_2$ and additional information on
the zero-energy ground-state scattering wave function from PAS
\cite{mms05,ykt06}. A multipolar van der Waals expansion in
$C_n/R^n$ is used to represent the potential at longer range
($R>22.7$\,$a_0$), and the gap between the two regions is bridged
with a spline interpolation to insure a smooth connection. The wave
functions are calculated using a full quantum calculation
\cite{mms05}.

As was the case in the analysis of \cite{skt08}, we do not have
enough information to independently determine all the van der Waals
coefficients, and improve on the precise relativistic many-body
calculation of \cite{pde06}, which gave $C_6=3103(7)$\, a.u.,
$C_8=3.792(8) \times 10^{5}$ a.u., and $C_{10}=4.215\times 10^7$
a.u. The last bound level $(v=62,J=0)$ is very extended, with its
outer turning point at roughly $R\sim 100$ $a_0$. At this point, the
leading contribution to the dispersion energy,
$-C_6/R^6-C_8/R^8-C_{10}/R^{10}$, arises from the $C_6$ term. In
fact, using the values of $C_n$ above, the $C_8$ contribution is
only 1.22\% of that of $C_6$, while the $C_{10}$ contribution
accounts for only 0.02\%. At a shorter distance $R\sim 20$ $a_0$,
similar to the maximum separation of states measured in
\cite{skt08}, these contributions are roughly 30.5\% and 8.5\%,
respectively. In \cite{skt08}, the precise theoretical values of
\cite{pde06} for $C_6$ and $C_8$ were used to fit the value of
$C_{10}$, since the energy levels $(v=0-50)$ were more deeply bound
and corresponded to shorter range than our levels $(v=62,J=0$ and
2). We note that less precise values of $C_6$ and $C_8$ from
\cite{mbr03} were also considered in \cite{skt08}.
%is a reevaluation of
%\cite{pde02} inputting
% an improved value of the $5s5p$\,$^1P_1$ lifetime \cite{ykt06}.
% It
%$C_6$ values reported in the
% literature are from a semiempirical method yielding
% $C_6=3250$\, a.u. \cite{mbr03}, and a
% relativistic many-body calculation %(typically good to a few percent)
% of $C_6=3170$\, a.u. \cite{pde02}.  Our result
%  is in reasonable agreement with \cite{pde06}.

If we use $C_6$ as a fit parameter to match the binding energy of
the $J=0$, $v=62$ state, assuming $C_8$ and $C_{10}$ from
\cite{pde06}, the best fit value is $C_6=3151(1)$\, a.u. where the
quoted uncertainty only reflects uncertainty in the measured binding
energy (see Table~\ref{tablebindingenergies}). The resulting
$^{88}$Sr X$^1\Sigma_g^+$ $s$-wave scattering length is $a=-2.0(2)$
$a_0$, where the uncertainty also only reflects uncertainty in the
measured binding energy. If we use the value of $C_{10}=6.60\times
10^{7}$ a.u. from \cite{skt08} instead, we find $C_6=3116.0(5)$\,
a.u., in reasonable agreement with \cite{pde06}. The resulting
scattering length is $a=-1.2(2)$ $a_0$. If instead, we take $C_6$
and $C_8$ from \cite{pde06}, and fit $C_{10}$, as was done in
\cite{skt08}, we find $C_{10}=7.488\times 10^7$ a.u. and $a=-0.9(2)$
$a_0$ (again, with the uncertainties reflecting the uncertainty in
the measured binding energy).

\begin{table}[!t]
\begin{tabular}{|l|c|c|c|c|c|}
 \hline
 Isotopes & 2-phot. & Fourier-& 1-phot. & 1-phot. & Thermal-  \\
         & PAS &   transform & PAS        & PAS &  ization  \\
  & [this study] & \cite{skt08}& \cite{mms05} & \cite{ykt06} & \cite{fdp06} \\
    & $a$ & $a$& $a$ & $a$ & $|a|$ \\
   \hline
 88-88 & -1.4(6) & $0(5)$ &$10^{+3}_{-11}$ & $-40^{+40}_{-100}$ & $21^{+3}_{-4}$\\
 \hline
87-87 & 96.2(1) & $97(2)$     &-& - & - \\
\hline
86-86 & 823(24) & $1050(380)$ &$1000^{+1300}_{-400}$ & - & $430^{+80}_{-90}$ \\
 \hline
84-84 & 122.7(3) &  $124(3)$    &-& - & - \\
 \hline
88-87 & 55.0(2) & $56(2)$     &-& - & - \\
\hline
88-86 & 97.4(1) & 99(2)& - & - & $110^{+10}_{-20}$  \\
 \hline
88-84 & 1790(130) & $>1170$ or    &-& - & - \\
      &        & $<-1900$     & &   &   \\
\hline
87-86 & 162.5(5) &  165.5(5.5)    &-& - & - \\
\hline
87-84 & -56(1) &  -55(10)    &-& - & - \\
\hline
86-84 & 31.9(3) &  $33(3)$    &-& - & - \\
  \hline
  \end{tabular}
\caption{Published values of scattering lengths given in units of
$a_0=0.053$\,nm}
  \label{tablescatteringlengths}
\end{table}

It is difficult to assess the uncertainties in $a$ related to these
coefficients. The most conservative assessment encompasses the full
range of values quoted here; $a=-1.4(6)$ and $C_6=3130(20)$\, a.u.
The uncertainties in $C_6$ and $C_8$ quoted in \cite{pde06} are
quite small, however, and no uncertainty is quoted for $C_{10}$. So
that might give more credence to the results for $C_6$ and $C_8$
from \cite{pde06} and the resulting fit $C_{10}=7.488\times 10^7$
a.u, which corresponds to the higher ends of the ranges of values
for $a$ in Table \ref{tablescatteringlengths}. Mass-scaling can be
used to determine the scattering lengths for all stable-isotope
collisional combinations from this information about the potential
(Table \ref{tablescatteringlengths}).

We note that the rotational energy takes the form
$H_{rot}=B(R)J(J+1)$ where $B(R)=\hbar^2/(2\mu R^2)$ is the
rotational constant for separation $R$ and reduced mass $\mu$. The
rotational constant for $v=62$ calculated using this potential
yields a binding energy of $-66.5$\,MHz for the $J=2$, $v=62$ state
(Table \ref{tablebindingenergies}), well within the measurement
uncertainty. We also found that this quantity does not significantly
constrain the $C_n$ coefficients.

 \begin{figure}
 %%%%%Use Screen size centered on page when exporting from matlab!!!
 %%THis saves the sizes%
%print to pdfwriter from ppoint,open resulting pdf with ghostview, convert to
%eps with pswrite, max resoltuion
  %\includegraphics[width=3in,clip=true,trim=100 290 50 80 ]{energydiagramstraight.eps}\\
  %\includegraphics[width=3in,clip=true]{PASdrawing.eps}\\
  %\includegraphics[width=2.25in,clip=true, trim=80 23 80 0, angle=270]{fig9.eps}%BosonicCrossSecVsEnergy.eps}\\ %{node86.eps}\\
  \includegraphics[width=3.3in,clip=true, trim=0 0 0 0, angle=0]{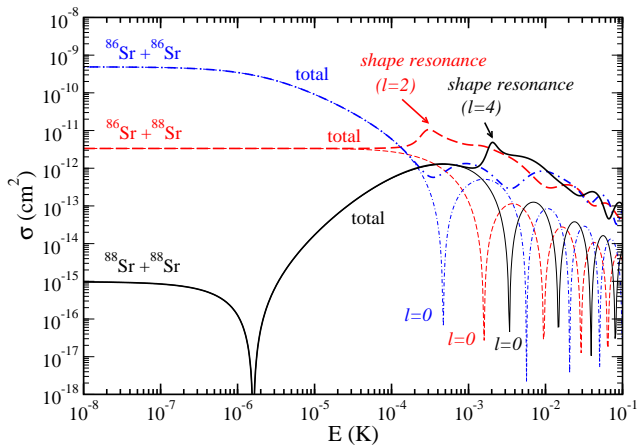}%BosonicCrossSecVsEnergy.eps}\\ %{node86.eps}\\
 \caption{Dependence of elastic-scattering cross sections on collision energy
  (E) in Kelvin
  for selected
  Sr isotopes (color online).  The thick lines  are cross sections
including partial waves up to $l=4$. Shape resonances are indicated.
Thin lines indicate cross section contributions from $l=0$ only. For
the plot, a potential is used that gives $a_{88-88}=-1.2\,a_0$ at
$E=0$. The cross sections are
 given by the usual expressions:
 %$\sigma = \frac{8\pi}{k^2}\sum_{\textrm{even}\, l = 0}^{\infty} (2 l +1)
% \sin^2 \delta_l (k)\stackrel{k\rightarrow 0}{\rightarrow} 8\pi a^2$
 $\sigma = ({8\pi}/{k^2})\sum_{l = 0,2,\ldots}^{\infty} (2 l +1)
 \sin^2 \delta_l (k)\stackrel{k\rightarrow 0}{\rightarrow} 8\pi a^2$
 for identical bosons
and $\sigma = ({4\pi}/{k^2})\sum_{l=0}^{\infty} (2 l +1) \sin^2
\delta_l (k)\stackrel{k \rightarrow 0}{\rightarrow} 4\pi a^2$ for
distinguishable atoms. The phase shift, $\delta_{l=0}(k)$, depends
on $k=\sqrt{2\mu E}/\hbar$ and is related to the scattering length,
$a$, and effective range, $r_e$, at low $k$ by $k\cot \delta_{l=0}
(k) = -\frac{1}{a} + \frac{1}{2} r_e k^2$.
  %All frequencies
%are measured with respect to the atomic $^1S_0$-$^1P_1$ transition.
  }\label{figure:BosonicCrossSecVsEnergy}
\end{figure}

In the ultracold regime, the energy dependence of scattering lengths
can often be neglected. However, this is not the case when there is
a low-energy scattering resonance or when the zero-energy scattering
length is very small. Figure \ref{figure:BosonicCrossSecVsEnergy}
demonstrates that
 $^{88}$Sr-$^{88}$Sr and $^{86}$Sr-$^{86}$Sr collision cross
 sections vary significantly with collision energy,
 even at energies below $1\,\mu$K.
 This may explain the discrepancy between this work and previous studies
 of Sr collisional properties \cite{fdp06}.

\section{Conclusion\label{Conclusion}}
%\subsection{}
%\subsubsection{}

Using two-photon photoassociative spectroscopy, we have measured the
binding energy of the $J=0$ and $J=2$ rotational levels of the
$v=62$ vibrational state of the X$^1\Sigma_g^+$ potential of
$^{88}$Sr$_2$. This is the least-bound ground vibrational level.
Combined with an accurate short range potential \cite{skt08} and
calculated van der Waals coefficients \cite{pde06}, the measurement
allows an accurate determination of $a$ for
 $^{88}$Sr-$^{88}$Sr interactions. Through mass scaling, we
determine the scattering lengths for all other isotopic
combinations. These measurements serve as a stringent test of atomic
structure calculations for alkaline-earth atoms and will provide
valuable input for future experiments with ultracold strontium.

We thank P. Julienne, R. Hulet, and E. Tiemann for helpful
discussions. P.P. was supported by the U.S. Department of Energy,
Office of Basic Energy Sciences. This work was supported by the
Welch Foundation (C-1579), National Science Foundation (PHY-0555639
and PHY 0653449), and the Keck Foundation.

%\bibliography{bibliography}

\end{document}